\documentclass[12pt]{elsarticle}

\usepackage[english]{babel}
\usepackage{hyperref}
\usepackage{comment}
\usepackage{soul}
\usepackage{color}
\usepackage{graphicx}
\usepackage{amsmath,amssymb}
\usepackage{soul}
\usepackage{lineno}

\begin{document}

\title{Probing short-range correlations in asymmetric nuclei
with quasi-free pair knockout reactions}

\author[SRC]{Sam Stevens}
\ead{Sam.Stevens@UGent.be}

\author[SRC]{Jan Ryckebusch}%
\ead{Jan.Ryckebusch@UGent.be}
\author[SRC]{Wim Cosyn}
\ead{Wim.Cosyn@UGent.be}

\author[W]{Andreas Waets}

\address[SRC]{%
 Department of Physics and Astronomy, Ghent University, Belgium
}%

\address[W]{Department of Physics and Astronomy, University of Leuven, Belgium}%


\date{\today}

\begin{abstract}
Short-range correlations (SRC) in asymmetric nuclei with an unusual neutron-to-proton ratio can be studied with quasi-free two-nucleon knockout processes following the collision between accelerated ions and a proton target.   
We derive an approximate factorized cross section for those SRC-driven $p(A,p^{\prime} N_1 N_2)$ reactions.
Our reaction model hinges on the factorization properties of SRC-driven $A(e, e^\prime N_1 N_2)$ reactions for which strong indications are found in theory-experiment comparisons. In order to put our model to the test we compare its predictions with results of $^{12}\text{C}(p,p^{\prime} pn)$ measurements conducted at Brookhaven National Laboratory (BNL) and find a fair agreement. The model can also reproduce characteristic features of SRC-driven two-nucleon knockout reactions, like back-to-back emission of the correlated nucleons.  We study the asymmetry dependence of nuclear SRC by providing predictions for the ratio of  proton-proton to proton-neutron knockout cross sections for the carbon isotopes $^{9-15}$C thereby covering neutron excess values  $(N-Z)/Z$ between -0.5 and +0.5.
\end{abstract}

\maketitle

\section{Introduction}
Many nuclear properties can be captured by the independent-particle model (IPM) that was developed in the fifties of the previous century
and has not lost its attractiveness as a predictive and descriptive nuclear  model  ever since. The nucleus, however, turns out to be more than the linear sum of its nucleons and a rich range of nuclear features fall beyond the scope of the IPM. For example, nuclear long-range correlations (LRCs), here loosely defined as correlations that extend over distance scales of the order of the nuclear radius, give rise to exciting collective phenomena like giant resonances and halo nuclei. The corresponding excitation-energy scale of nuclear LRCs is well established and is of the order of MeVs. Nuclear short-range correlations (SRCs) \cite{Arrington:2011xs,Atti:2015eda,Hen:2016kwk} 
extend over distance scales of the order of the nucleon size, and are connected with substantially larger energy-momentum scales than the LRCs.  

The scale separation between IPM, LRC and SRC effects is manifested in nuclear 
momentum distributions $n_A(k)$. The IPM can account for the strength below the 
Fermi momentum $k_F \approx$220~MeV/c. The impact of LRCs on the $n_A(k)$ at 
high nucleon momenta is mainly confined to $k \gtrsim k_F$ 
\cite{Dickhoff:2004xx} and the tensor component of the SRC is the major 
source of strength for the fat 
tails above the Fermi momentum~\cite{Hen:2014nza}. This has important 
implications, for example, for the second moment $\left< k ^2 \right>$ of 
$n_A(k)$, that can be connected with the expectation value of the 
non-relativistic kinetic energy $\left< \frac {k ^2} {2m_N} \right> $. Indeed, 
in an IPM the majority component (most often neutrons) has a larger expectation 
value for the second moment $\left< k ^2 \right>$ than the minority component 
(most often protons). The dominant role of the tensor component in the SRC 
turns this picture upside down and provides the 
substantially larger values for the $\left< \frac {k ^2} {2m_N} \right> $ of 
the minority component (protons) 
\cite{Sargsian:2012sm,Wiringa:2013ala,Vanhalst:2014cqa}.  This illustrates that 
there are important $(N-Z)$ asymmetry aspects to SRCs that are awaiting further 
explorations.
 
While being a fascinating phenomenon in itself, a full understanding of nuclear SRCs and its $(N-Z)$ asymmetry dependence is pivotal for studies of compact objects like neutron stars and of the nuclear equation of state~\cite{Fortin:2016hny,Oertel:2016bki}.  The last couple of decades have marked significant growth in insight into the mass and isospin dependence of nuclear SRCs thanks to an experimental program of quasi-free two-nucleon knockout reactions with hadronic and electroweak probes \cite{Onderwater:1997zz,Giusti:1997pa,Onderwater:1998zz,ROSNER200099,Starink:2000qhh,Tang:2002ww,Shneor:2007tu,Korover:2014dma,Colle:2015ena}. Measuring the multi-fold cross sections for 
those reactions with at least four particles in the final state, is really challenging which made one to think about alternate ways of addressing SRC physics through nuclear reactions. Substantial progress in extracting the SRC physics from $A(e,e^\prime N_1N_2)$ measurements has been recently made by investigating appropriate ratios of aggregated cross sections and observables against well-selected variables \cite{Hen:2014nza, Colle:2015ena, Ryckebusch:1996wc, Colle:2015lyl}. Thereby, proper kinematic cuts and selections have  been key to success.   
Strong evidence for the proton-neutron dominance of SRC, for example, emerges from comparing measured $A(e,e'pp)/A(e,e'p)$ and $A(e,e'pn)/A(e,e'p)$ ratios \cite{Hen:2014nza}. The mass dependence of SRC could be addressed with the aid of measured $A(e,e'pp)/^{12}\text{C}(e,e'pp)$ and $A(e,e'pn)/^{12}\text{C}(e,e'pn)$ ratio's~\cite{Colle:2015ena}. The results of these studies provided strong evidence that for mid-heavy and heavy nuclei the aggregated impact of SRC roughly scales with the mass number ($\sim A$) and not with $\sim A^2$ as naively expected.   

A remaining question, particularly important for understanding the physics of neutron stars for example, is how the SRC evolve with the asymmetry $(N-Z)$ between the number of neutrons and protons. This question can be addressed with selected reactions at radioactive-beam facilities. Not only can one probe nuclei with an exotic $(N-Z)$, the possibility to measure also the properties of the remnant nucleus, adds an additional layer of accuracy and detail that has not been achieved in most of the $A(e,e'N_1N_2)$ studies addressing SRC.   

Due to new advanced techniques and equipment, the potential of quasi-free $(p,p^{\prime}p)$ single-nucleon knockout in inverse kinematics has been demonstrated to provide  the means to study single-particle properties for short-lived nuclei  \cite{Aumann20073,Kobayashi:2008zzc,Panin:2016div}. Along the same lines, the development of a program for $(p,p^{\prime}N_1N_2)$ SRC studies from short-lived nuclei with energies of the order of GeV per nucleon is discussed in the community. The success of studies using nucleon knockout reactions in quasi-free kinematics very much hinges on the availability of an approximate expression for the multi-fold cross sections. For example, the quasi-free hypothesis \cite{Jacob:1973tu,Kitching:1984gc} has proven its great value and effectiveness in studies of the single-particle properties of nuclei with 
$(p,p^{\prime}p)$ reactions. The quasi-free hypothesis requires that the transferred energy is large compared to the average binding energy and that the ejected nucleon $N$ carries away most if not all of the transferred energy and momentum.   The availability of a factorized approximate form for the cross sections, is particularly important in the planning phase of the experiments. For example, it is pivotal in order to get realistic estimates of the expected count rates. The main purpose of this paper is to provide a factorized expression for the SRC driven quasi-free $p(A,p^\prime N_1 N_2 R^\ast)$ reaction. We build our reaction model on a formalism that resulted in a factorized expression for exclusive $A(e,e^{\prime}N_1N_2)R^\ast$ reactions that has been well tested in several theory-experiment comparisons  \cite{Colle:2015ena, Ryckebusch:1996wc,Colle:2015lyl, Ryckebusch:1997gn, Blomqvist:1998gq, Colle:2013nna}. Thereby we establish a connection between the correlation functions that account for nuclear SRC effects in the nuclear momentum distributions $n_A(k)$ and the two-nucleon knockout cross sections.

\section{Formalism}
\label{sed:formalism}
\begin{figure}[ht]
\centering
\includegraphics[width=\textwidth]
{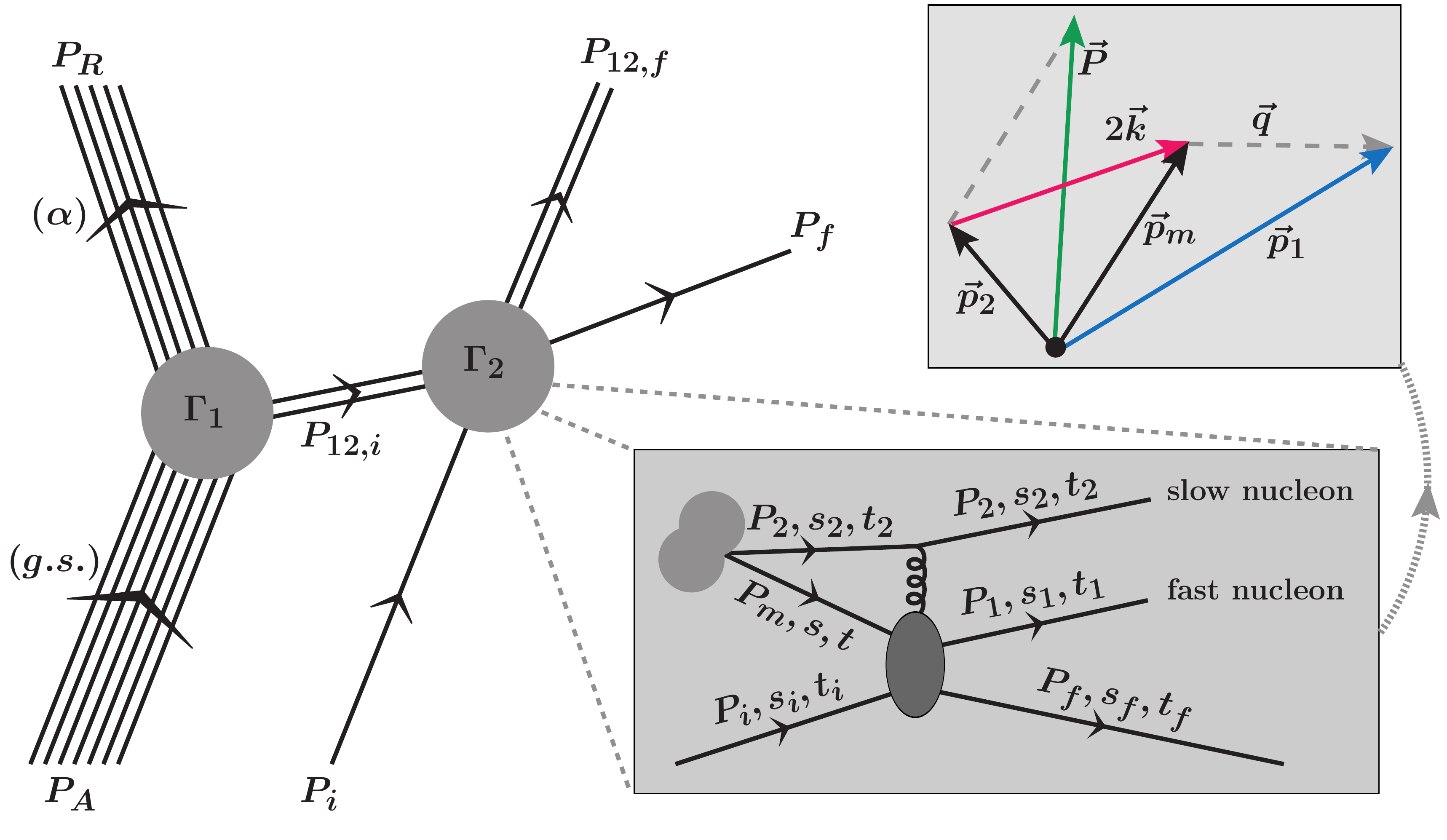} 
\caption{Pictorial diagram of a kinematically complete $p \left( A,p^\prime N_1 
N_2 R ^\ast \right)$ reaction in quasi-free kinematics.  The reaction is induced 
by an accelerated ion $A$ (four-momentum $P_A$) in its ground state colliding 
with a target proton ($P_i$). As a result, an SRC nucleon-nucleon (NN) pair gets 
ejected from the ion leaving a residual fragment $R^{\ast}$ ($P_R^\ast$). 
The initial and final four-momentum  of the SRC NN pair are $P_{12,i}$ and 
$P_{12,f}$.
In the final state the pair consists of two asymptotically free nucleons with 
four-momenta $P_1(E_1, \vec{p}_1)$ and $P_2(E_2, \vec{p}_2)$. The recoiling 
target proton has four-momentum $P_f$. The residual nucleus $R^\ast$ is created 
in a state designated by the quantum numbers $\alpha$.
The shaded boxes illustrate the momentum variables (top) and the spin-isospin   
$(s,t)$ variables (bottom) used to describe the collision between an SRC NN pair 
and a proton in the IA.} 
\label{figPPNNreaction}
\end{figure}

We now develop an approximate but realistic framework to compute  cross sections 
for $p(A,p^\prime N_1 N_2 R^\ast)$ reactions (see Fig.~\ref{figPPNNreaction} for 
momenta and spin-isospin labels). Our model applies to reactions where the 
accelerated ion's energy is sufficiently large to adopt the quasi-free 
hypothesis \cite{Jacob:1973tu, Kitching:1984gc} that justifies both the impulse 
approximation (IA) and the spectator approximation (SA). The IA implies that the 
target proton interacts with a single nucleon in the SRC NN pair.  In the SA 
only the correlated pair in the accelerated ion gets directly involved in the 
proton collision process and all other nucleons act as 
spectators~\cite{Stetz:1980zz}. The quasi-free reaction picture for 
$p(A,p^\prime N_1 N_2 R^\ast)$ is schematically shown 
in~Fig.~\ref{figPPNNreaction}. As is commonly the case with quasi-free 
processes, the transition matrix element $\mathcal{M}$ can be approximately 
factored in a nuclear-structure $ \left( \mathcal{M}_1 \right)$ and a 
nuclear-reaction $ \left( \mathcal{M}_2 \right)$ part 
\begin{align}
\label{eqTransitionMatrixElementFactorisation}
& \mathcal{M}(P_i + P_A \rightarrow P_f + P_{12,f} + P_{R}^\ast) 
\approx
\nonumber \\
& \qquad 
\mathcal{M}_1
\left( \Gamma_1; \, P_A \rightarrow P_{R}^\ast + P_{12,i} \right) \;
\mathcal{M}_2 
\left( \Gamma_2; \, P_{12,i} + P_i \rightarrow P_{12,f} + P_f \right) \, .
\end{align}
The vertex $\Gamma_1$ encodes at given kinematics the probability of removing a 
bound SRC NN pair from the accelerated ion. Reaction vertex $\Gamma_2$ describes 
proton scattering from an SRC NN pair resulting in three asymptotically free 
nucleons. For the calculation of the amplitude of the $\Gamma_2$ vertex (see 
inset of Fig.~\ref{figPPNNreaction}) we build on our derivations of a factorized 
cross section for electro- and photo-induced SRC NN pair knockout reactions 
\cite{Colle:2015ena,Ryckebusch:1996wc,Colle:2015lyl,Ryckebusch:1997gn,
Colle:2013nna}. Thereby, the transferred momentum $\vec{q} = \vec{p}_i - 
\vec{p}_f$ is fully absorbed by a single nucleon (the so-called ``fast 
nucleon'') in the SRC NN pair. The other nucleon in the SRC NN pair is referred 
to as the ``slow nucleon'' and is ejected as a result of the ``broken'' 
correlation. In the projectile frame (PF), that is the rest 
frame of the accelerated ion $A$, the missing energy $E_m$ and missing momentum 
$\vec{p}_m$ in the 
$\Gamma_2$ interaction are defined as 
\begin{align}
\label{eqMissingMomentum}
E_m = \left(E_1 + E_f - E_i - m_{N_1}\right)_{\text{PF}} \qquad ,
\qquad \vec{p}_m = \left(\vec{p}_1 + \vec{p}_f - \vec{p}_i\right)_{\text{PF}} \, ,
\end{align}
where $m_{N_1}$ is the mass of the fast nucleon. In the IA, $\vec{p}_m$ 
corresponds with the momentum of the nucleon that scatters with the target 
proton.  Ignoring medium modifications, the collision between the fast nucleon 
in the SRC NN pair and the proton target can be modeled as free nucleon-proton 
scattering  
\begin{align}
\label{eqFreeProtonProtonInteraction}
\left(P_m \equiv \left(E_m + m_{N_1}, \vec{p}_m \right), s, t\right) + (P_i, 
s_i, t_i) \rightarrow (P_1, s_1, t_1) + (P_f, s_f, t_f) \, ,
\end{align}
with $s_j$ and $t_j$  the spin and isospin projections of the nucleon $j$.

\begin{figure}[h]
\centering
\includegraphics[width=\textwidth]{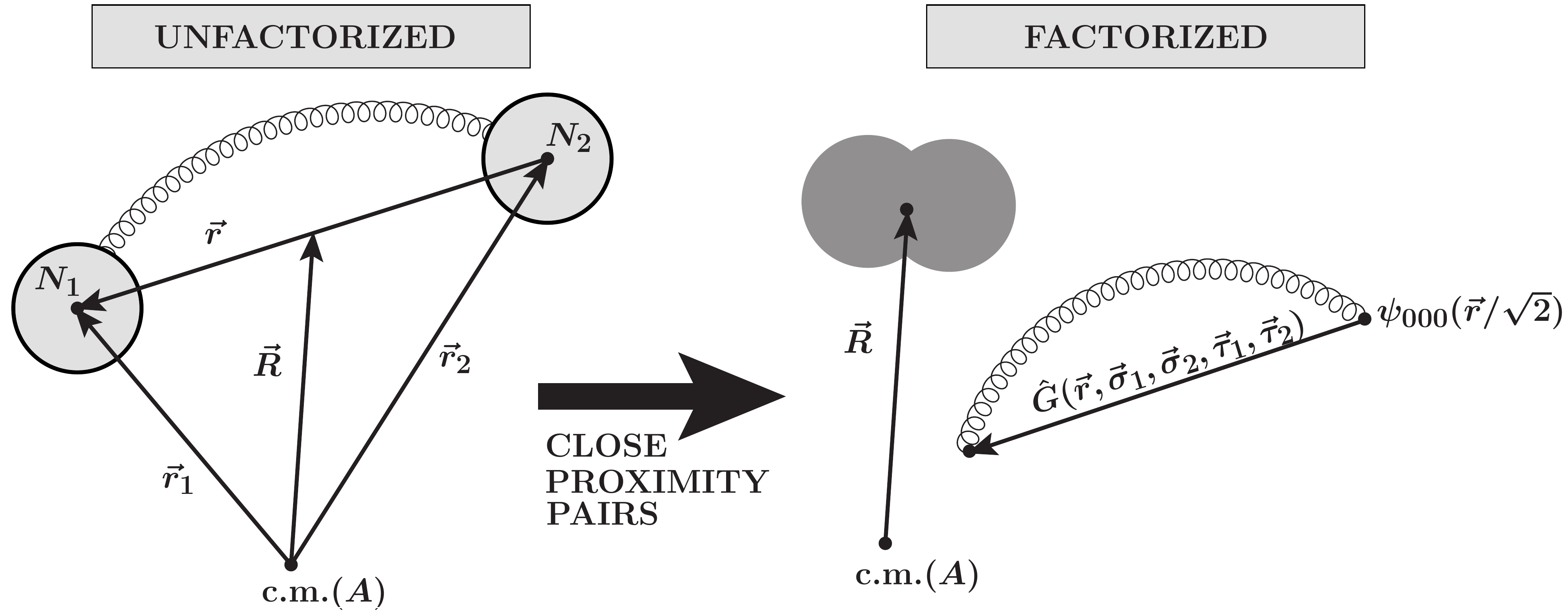} 
\caption{Schematic representation of the spatial coordinates of a correlated NN 
pair. After projecting on close-proximity pairs, the dependence on the c.m.~and 
relative coordinates can be factorized.}
\label{figRelcomcoord}
\end{figure}

Given the possibility of detecting all final-state fragments in radioactive beam 
experiments, we compute the amplitude for given angular momentum and isospin of 
the  ion (=$J_A T_A$) and of the residual fragment (=$J_R T_R$) under conditions 
that allow one to identify the slow nucleon and the isospins of all 
asymptotically free nucleons. The transition matrix element of 
Eq.~(\ref{eqTransitionMatrixElementFactorisation}) involves a summation over all 
allowed $J, \; T$ combinations of the SRC NN pair and is evaluated in the 
PF
\begin{align}
\label{eqM}
& \mathcal{M}
 =
 \frac{1}{\sqrt{2}}
 \sum_{\substack{J M T M_T}}
 \sqrt{\mathcal{Z}(\Gamma_1)} \;
 \left\langle 
 J_R \, M_R \, J \, M
 \big\vert
 J_A \, M_A
 \right\rangle \,
 \left\langle 
 T_R \, M_{T,R} \, T \, M_T
 \big\vert
 T_A \, M_{T,A}
 \right\rangle 
 \nonumber \\
& \qquad \times
 \Big\langle \frac{1}{2} \, t_1 \, \frac{1}{2} \, t_2
 \Big\vert T \, M_T \Big\rangle 
 \sqrt{\frac{E_2m_{N_1}}{(E_m + m_{N_1})m_{N_2}}}
 \sum_{s} M^{pN_1}_{NN\to NN}
\nonumber \\ & \qquad \times
 \frac{1}{(2 \pi)^3} 
 \int \text{d} \vec{R} \;\text{d} \vec{r} \;
 e^{-i \vec{P} \cdot \vec{R}} \,
 e^{-i \vec{k} \cdot \vec{r}}
 \left\langle
 \vec{R} \, , \, \vec{r} \, ; \, s \, t_1 \, , \, s_2 \, t_2
 \Big\vert
 (\beta \gamma) 
 J M T M_T
 \right\rangle_{SRC}
 \, ,
\end{align}
with
\begin{align}
\mathcal{Z}(\Gamma_1) \equiv \mathcal{Z}(\Gamma_1; J_R T_R, J T)
=
 \frac{A(A-2)}{\mathcal{N}_{A, \text{corr}} \, \mathcal{N}_{R, \text{corr}}} 
\frac{\mathcal{S}(J_R T_R, J T, \beta \gamma)}{N_J N_T N_{J_R} N_{T_R}} \; .
\end{align}
Here, $\mathcal{S}(J_R T_R, J T, \beta \gamma)$ is a spectroscopic factor, 
$N_J$, $N_T$, $N_{J_R}$ and $N_{T_R}$ denote the number of possible quantum 
numbers for the SRC NN pair, and $\mathcal{N}_{A, \text{corr}}$ and 
$\mathcal{N}_{R, \text{corr}}$ are normalization factors. The first three 
factors in the above expression for the amplitude $\mathcal{M}$ find their 
origin in the vertex $\Gamma_1$. In the derivation of
Eq.~(\ref{eqM}),the normalization of states is treated relativistically, 
whereas the dynamics is treated non-relativistically through wave functions 
evaluated in the PF.
The initial relative and c.m.~momenta  of the SRC NN pair (see also 
Fig.~\ref{eqTransitionMatrixElementFactorisation}) are defined as
\begin{align}
\label{eqInitialMomenta}
\vec{P} \equiv \vec{p}_{12,i}^{\;\text{cm}} = \vec{p}_m + \vec{p}_2
\qquad \qquad
\vec{k} \equiv \vec{p}_{12,i}^{\;\text{rel}} = \frac{\vec{p}_m - \vec{p}_2}{2} 
\, .
\end{align}
The corresponding conjugated c.m.~and relative coordinates read 
\begin{align}
\label{eqComRelCoord}
\vec{R} = \frac{\vec{r}_1 + \vec{r}_2}{2}
\qquad \qquad
\vec{r} = \vec{r}_1 - \vec{r}_2 \, ,
\end{align}
with $\vec{r}_1$ and $\vec{r}_2$ (see Fig.~\ref{figRelcomcoord}) the spatial 
coordinates of the SRC NN pair in the projectile frame.

In the expression of Eq.~(\ref{eqM}), $M^{pN_1}_{NN\to NN}$ is the matrix 
element for elastic nucleon-nucleon scattering and the $\left\vert (\beta 
\gamma) J M T M_T \right\rangle_{SRC}$ determines the quantum state of the SRC 
NN pair. It is     
constructed as in Ref.~\cite{Colle:2013nna} that contains a derivation of a 
factorized cross section of SRC-driven $A(e,e'N_1N_2)R^{\ast}$. 
The SRC pair's state is modeled as a correlation operator acting on a normalized 
and anti-symmetric (nas) state of two IPM nucleons characterized by the quantum 
numbers $\beta = \left(n_{\beta}, l _{\beta}, j_{\beta}, t_{\beta} \right)$ and 
$\gamma = \left(n_{\gamma}, l _{\gamma}, j_{\gamma}, t_{\gamma} \right)$ 
respectively. As a result, the residual fragment's state $\alpha$ can be 
specified by a two-hole state $\left\vert \beta^{-1} \gamma^{-1} \right\rangle$ 
in the ground state of the initial nucleus.  
With these assumptions, one can write for the state of the SRC NN pair 
\begin{align}
\label{eqCorrelatedCoupledState}
\left\vert (\beta \gamma) J M T M_{T} \right\rangle_{SRC}
=
\widehat{G}
(\vec{r}, \vec{\sigma}_1, \vec{\sigma}_2, \vec{\tau}_1, \vec{\tau}_2) \,
\left\vert n_\beta n_\gamma l_\beta l_\gamma (j_\beta j_\gamma) J M (t_\beta 
t_\gamma) T M_T \right\rangle_\text{nas} \, .
\end{align}

We use harmonic oscillator (HO) single-particle states as they offer the 
possibility to separate the pair's relative and c.m.~motion with the aid of 
Moshinsky brackets 
$\left\langle \ldots \right\rangle_\textrm{Mos}$. 
It has been numerically shown \cite{Vanhalst:2014cqa, Colle:2015ena,  
Ryckebusch:1996wc, 
Colle:2015lyl, Ryckebusch:1997gn, Colle:2013nna} that in evaluating expressions 
of the type (\ref{eqCorrelatedCoupledState}) the major source of SRC strength 
stems from correlation operators acting on IPM pairs with relative quantum numbers 
$(n=0 ,\; l=0)$. This can be intuitively understood by noting that the 
probability of finding close-proximity IPM pairs is dominated by pairs in a 
relative $(n=0, \; l=0)$ state (see also Fig.~\ref{figRelcomcoord}). In 
Ref.~\cite{Vanhalst:2014cqa} it was shown that across the nuclear chart 
about 90\% of the fat tail in the $n_A(k)$ finds its origin in correlation 
operators acting on IPM pairs in a relative $(n=0, \; l=0)$ state.   

This leads to the following approximate expression for the wave function of the 
SRC NN pair in coordinate space:
\begin{align}
\label{eqCorrelatedCloseProximityPair}
&  \left\langle
 \vec{R} , \vec{r} \,
 \Big\vert
 (\beta \gamma) 
 J M T M_T
 \right\rangle_{SRC}
 \approx
 2
 \sum_{N L M_L M_S}
 \mathcal{C}^{\beta \gamma}_{00JM}
 \left( 
    S, M_S, N, M_L
 \right) \psi_{N L M_L}(\sqrt{2} \vec{R})
 \nonumber \\
& \quad \times
 \psi_{000}\bigg(\frac{\vec{r}}{\sqrt{2}}\bigg) \;
 \widehat{G}
 (\vec{r}, \vec{\sigma}_1, \vec{\sigma}_2, \vec{\tau}_1, \vec{\tau}_2)
 \left\vert
 (1-T) \, M_S \, , \, T \, M_T
 \right\rangle \, ,
\end{align}
where $\psi_{n l m_l}\big(\frac{\vec{r}}{\sqrt{2}}\big)$ $ \left( \psi_{N L 
M_L}\big(\sqrt{2}\vec{R}\big) \right)$ are HO eigenstates for the pair's 
relative (c.m.)~motion and the following factor has been introduced  
\begin{align}
\label{eqC00}
& \mathcal{C}^{\beta \gamma}_{00JM}
 \left( 
    S, M_S, N, M_L
 \right)
\nonumber \\ 
& =
 \left\langle
 0 0, N (\varepsilon - 2N), (\varepsilon - 2N)
 \bigg\vert 
 n_\beta l_\beta, n_\gamma l_\gamma, (\varepsilon - 2N)
 \right\rangle_\textrm{Mos}
\nonumber \\
&  \times
 \sum_{s_b, s_c}
 \sum_{m_b, m_c}
 \sum_{m_\beta m_\gamma}
 \left\langle 
	j_\beta m_\beta j_\gamma m_\gamma
 \big\vert
 	J M
 \right\rangle
 \left\langle
 l_\beta m_b \frac{1}{2} s_b \bigg\vert j_\beta m_\beta
 \right\rangle
 \left\langle
 l_\gamma m_c \frac{1}{2} s_c \bigg\vert j_\gamma m_\gamma
 \right\rangle
\nonumber \\
& 
\times
 \left\langle 
 \frac{1}{2} s_b \frac{1}{2} s_c
 \bigg\vert
 (1-T) M_S
 \right\rangle
 \left\langle
 l_\beta m_b l_\gamma m_c 
 \big\vert
 (\varepsilon - 2N) M_L
 \right\rangle \, ,
\end{align}
with $\varepsilon = 2 n_\beta + l_\beta + 2 n_\gamma + l_\gamma$ the HO energy 
of the NN pair.

With respect to the correlation operator $\widehat{G}$, the central $(c)$, 
tensor $(t\tau)$, and spin-isospin $(\sigma \tau)$ terms are responsible for the 
majority of the nuclear SRC \cite{Vanhalst:2014cqa, Ryckebusch:1996wc, 
Colle:2013nna, 
Guardiola:1996dq,Benhar:1993zz,Pieper:1992gr,CiofidegliAtti:1999zz, 
Vanhalst:2012ur}
\begin{align}
\label{eqCorrelationOperator}
\widehat{G}
 		\bigl( \vec{r}, \vec{\sigma}_1, \vec{\sigma}_2, 
        \vec{\tau}_1, \vec{\tau}_2 \bigr)
 = 1 - f_c(r) + f_{t \tau}(r) \widehat{S}_{12} (\vec{\tau}_1 \cdot \vec{\tau}_2)
 + f_{\sigma \tau}(r) (\vec{\sigma}_1 \cdot \vec{\sigma}_2) (\vec{\tau}_1 \cdot 
\vec{\tau}_2) \, ,
\end{align}
with $\widehat{S}_{12}$ the tensor operator. The functions $f_c$, $f_{t \tau}$ 
and $f_{\sigma \tau}$ are the central, tensor and spin-isospin correlation 
functions. They are responsible for the fat tails in the nuclear momentum 
distributions and determine the SRC interaction strength at any given 
spin-isospin combination of the NN pair.   All results contained in this paper 
are obtained with a set of correlation functions that we have systematically 
used and tuned in SRC-driven reaction studies \cite{Onderwater:1998zz} 
\cite{Starink:2000qhh} \cite{Colle:2015ena} \cite{Blomqvist:1998gq}. The $f_{t 
\tau}(r)$ and $f_{\sigma \tau}(r)$ correlation functions are from a variational 
calculation~\cite{Pieper:1992gr}, the central correlation function $f_c(r)$ from 
the G-matrix calculations in nuclear matter~\cite{Gearheart1994}.

After averaging over initial and summing over final polarization states, the 
combination of 
Eqs.~(\ref{eqM}), 
(\ref{eqCorrelatedCloseProximityPair}) and (\ref{eqCorrelationOperator}),  leads 
to the following factorized differential cross section in the laboratory 
frame \\
\begin{align}
\label{eqCrossSection}
 \frac{\text{d} \sigma^{(pN_1N_2)}}
 {\text{d} \Omega_f \,
 \text{d} E_1 \, \text{d} \Omega_1 \,
 \text{d} E_2 \, \text{d} \Omega_2}
&=
 2^{\left\vert M_T \right\vert} \;
 \mathcal{K} \;
 \frac{\text{d} \sigma^{pN_1}}{\text{d} t}  \;\nonumber\\
 &\times\left\lbrace
 \frac{E_2}{E_m + m_{N_1}} \;
 \sum_{J \; M \; T}
 \frac{\mathcal{Z}(\Gamma_1; J_R T_R, J T)}
 {(2 J + 1)(2 T + 1)}
 F^{\beta \gamma}_{JM,T} (\vec{P}, \vec{k})
 \right\rbrace_\text{PF}
\end{align}
with $\mathcal{K}$ a kinematic factor evaluated in the laboratory frame
\begin{align}
 \mathcal{K} \, 
=
 \frac{1}{(2 \pi)^8} \;
 \frac{(P_f \cdot P_1)^2 - m_p^2 m_{N_1}^2}
 {\sqrt{\left( P_i \cdot P_A \right)^2 - m_p^2 m_A^2}} \;
 m_A m_{R^\ast} \;
 \frac{p_f p_1 p_2}{E_R} \;
 \left\vert
   1
   -
   \frac{E_f}{E_{R}}
   \frac{\vec{p}_R \cdot \vec{p}_f}{p_f^2}
 \right\vert^{-1} \; ,
\end{align}
$\frac{\text{d} \sigma^{pN_1}}{\text{d} t}$ (with $t \equiv (P_i - P_f)^2$)  
the cross section for free proton-nucleon scattering, determined at the 
off-shell kinematic invariants of the subprocess of  
Eq.~(\ref{eqFreeProtonProtonInteraction}), and the factor between curly 
brackets is evaluated in the projectile frame. The function $F_{J M, T}^{\beta 
\gamma} (\vec{P}, \vec{k})$ accounts for the SRC effects and can be factored 
into parts depending on the c.m.($\vec{P}$) and 
relative ($\vec{k}$) momentum of the SRC pair (see also 
Fig.~\ref{figRelcomcoord}) 
\begin{align}
& 
 F^{\beta \gamma}_{J M, T} (\vec{P}, \vec{k})
 = \sum_{\mu = T - 1}^{1 - T}
 \Bigg\vert
 \mathcal{F}^{(0)}_{\nu} [f_c - 3 f_{\sigma \tau}] (k) \;
 \mathcal{P}^{\varepsilon \beta \gamma}_{J M T \mu} (\vec{P}) 
-
 \delta_{T,0} \;
 12 \sqrt{2 \pi} \;
 \mathcal{F}^{(2)}_\nu [f_{t \tau}] (k)
 \nonumber \\
& \qquad \qquad \;\times
 \sum_{m_l = - 2}^{2}
 \left\langle
 2 m_l 1 \mu
 \big\vert
 1 (m_l + \mu)
 \right\rangle
 \mathcal{P}^{\varepsilon \beta \gamma}_{J M T (m_l + \mu)} (\vec{P}) \;
 Y_{2, m_l} (\Omega_k) 
 \Bigg\vert^2 \, .
\label{eq:Ffunction}
\end{align}
Obviously, the relative-momentum part receives contributions from all three 
terms in the correlation operator of Eq.~(\ref{eqCorrelationOperator}). At given 
$k$ the strength attributed to the correlation function $f$ is given by
\begin{align}
\label{eqLinearIntegralTransformation}
 \mathcal{F}^{(l^\prime)}  [f] (k)
& =
 \frac{4 \pi}{\sqrt{2l^\prime + 1}}  
 \left[
 \sum_{m_l^\prime= -l^\prime}^{l^\prime}
 \left\vert \,
    \int \frac {\text{d} \vec{r}} {(2 \pi)^{3/2}}  \, e^{-i \vec{k} \cdot 
\vec{r}} \,
    \psi_{000}\bigg(\frac{\vec{r}}{\sqrt{2}}\bigg) \,
    Y_{l^\prime m_l^\prime} (\Omega) \,
    f(r) \, 
 \right\vert^2 \,
 \right]^{1/2} \, .
\end{align}
At given c.m. momentum $\vec{P}$ the contribution to the $F^{\beta \gamma}_{J M 
T} (\vec{P}, \vec{k}) $ is determined by
\begin{align}
 \mathcal{P}^{\beta \gamma}_{J M T \mu} (\vec{P})
=
 \sum_N
 \mathcal{C}^{\beta \gamma}_{00JM}
 \left( 
    1-T, \mu, N, M - \mu
 \right)
 \int \frac{\text{d} \vec{R}}{(2 \pi)^{3/2}} \; 
 e^{- i \vec{P} \cdot \vec{R}} \, 
 \psi_{\substack{N L=(\varepsilon - 2N)\\ M_L=(M - \mu)}} 
\big(\sqrt{2}\vec{R}\big) \, .
\label{eq:roundPfunction}
\end{align}
\section{Results}
\label{sec:results}

All results presented here 
use the cross-section form of Eq.~(\ref{eqCrossSection}) based on plane-wave 
dynamics of the impinging and ejected nucleons ignoring initial- and final-state 
interactions (IFSI). The FSI analysis of SRC-driven $A(e,e^{\prime}N_1N_2)$ 
in~\cite{Colle:2015lyl} indicate that FSI mainly cause a reduction of the cross 
sections, without significantly changing their shape. For the cross sections 
with carbon beams we deduce an IFSI reduction factor of the plane-wave based 
cross sections of the order $0.05 - 0.1$. In what follows we present results for 
SRC driven 2N knockout from the $^{9-16}\text{C}$ isotopes. The HO 
single-particle states of those nuclei are obtained from an analysis of the 
momentum distributions extracted from the $p(^{9-16}\text{C},p^{\prime}p)$ 
measurements of Ref.~\cite{Kobayashi:2008zzc}. The values for the free 
proton-proton cross section 
$\frac{\text{d} \sigma^{pp}}{\text{d} t}$ in Eq.~(\ref{eqCrossSection}) are 
obtained from the SAID code \cite{Arndt:2000xc,ScatteringSAID} for laboratory 
kinetic energies below $3$ GeV and from the parametrization of 
Ref.~\cite{Uzhinsky:2016uhg} for higher energies.
 
\begin{figure}[!htb]
\centering
\includegraphics[width=\textwidth]{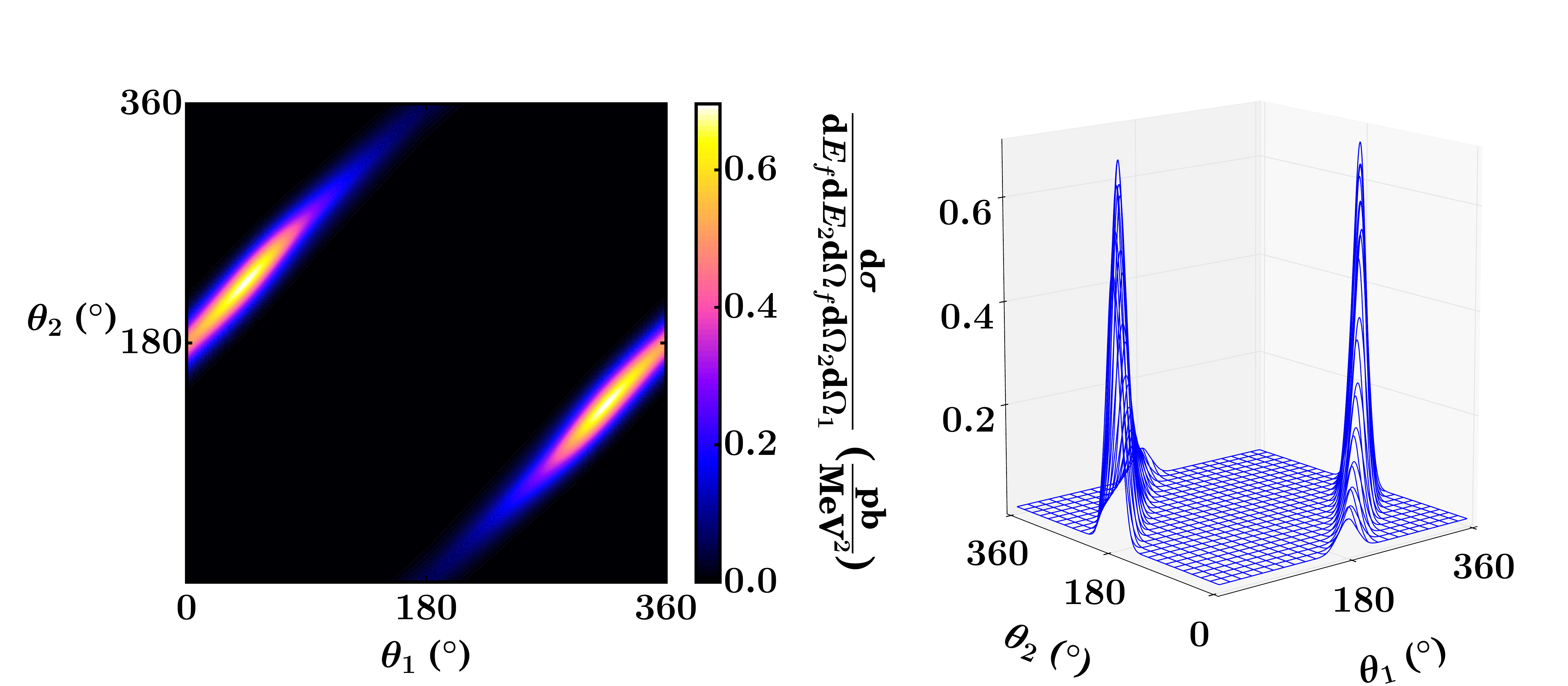} 
\caption{The $p({}^{\,10}\text{C},p^{\prime}pn)$ cross section for the knockout 
of an SRC $pn$ pair as a function of the bound proton's (neutron's) PF angle 
$\theta_1$ ($\theta_2$) relative to the initial proton momentum $\vec{p}_i$. All 
particle momenta lie in the ion scattering plane, with laboratory momenta $p_A = 
2.5$~A~GeV and $p_f = 1.5$~GeV. The nucleons' initial momenta in the PF are $p_m 
= p_2 = 400$~MeV.}
\label{figCrossSectionspppsss}
\end{figure}

Fig.~\ref{figCrossSectionspppsss} displays the 
$p({}^{\,10}\text{C},p^{\prime}pn)$ cross section in  specific in-plane 
kinematics. 
Clearly, the bulk of the cross sections comes from $pn$ knockout in 
configurations approaching back-to-back  of the initial SRC pair. This feature 
emerges in several SRC investigations ~\cite{Hen:2014nza, Colle:2015ena, 
Colle:2015lyl,  VanCuyck:2016fab}.

\begin{figure}
\centering
\includegraphics[width=0.9\linewidth]{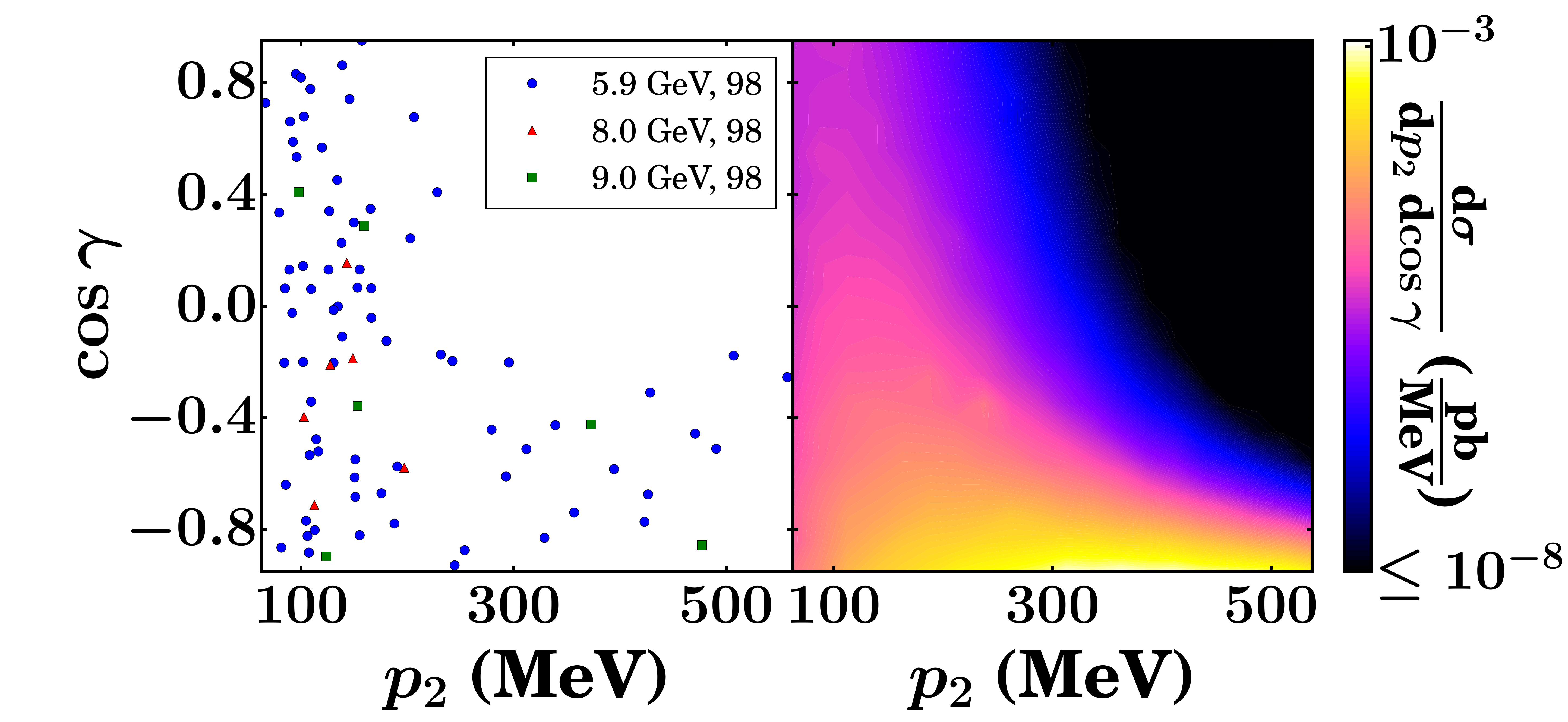}
\includegraphics[width=0.9\linewidth]{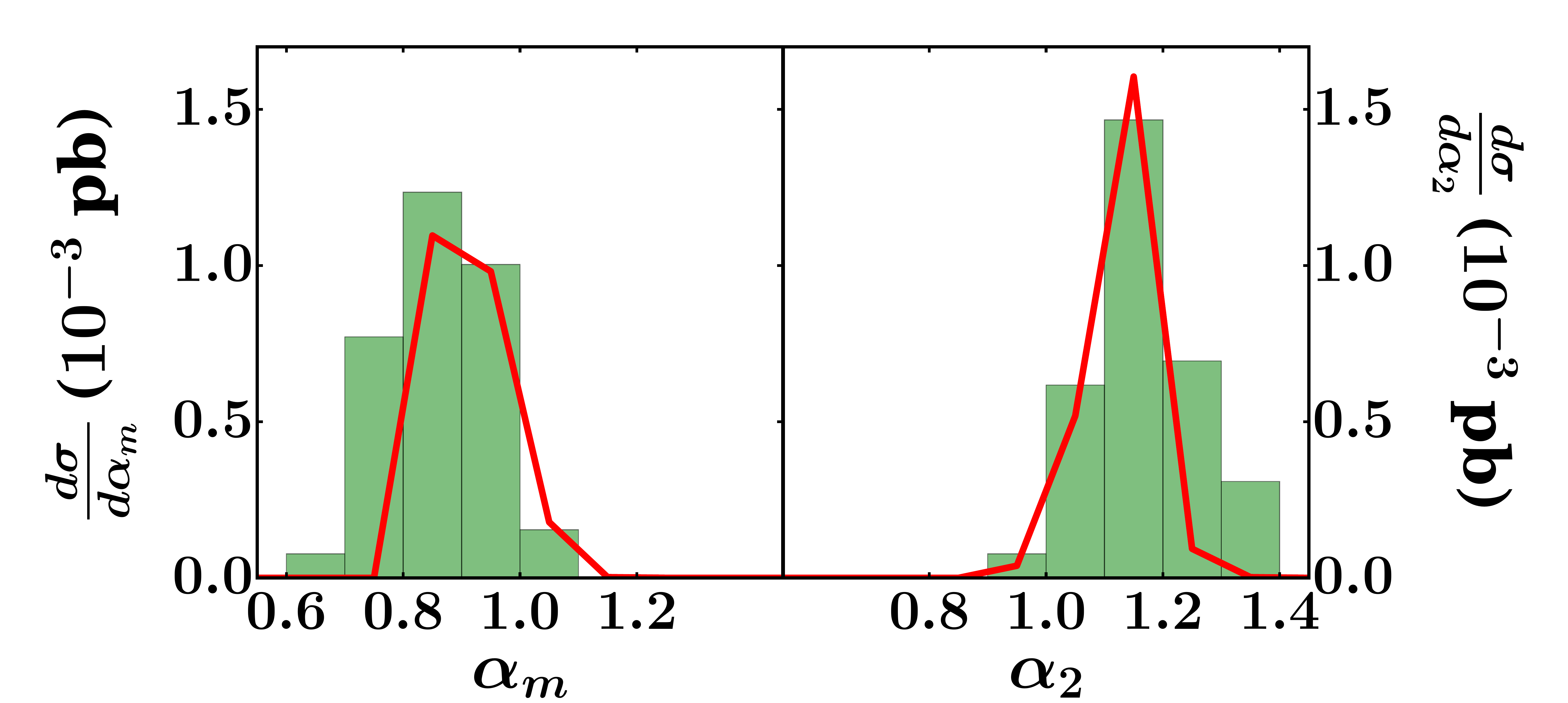} 
\caption{
Predictions for ${}^{\,12}\text{C}(p,p^{\prime}pn)$ observables in the 
kinematics of Ref.~\cite{Tang:2002ww}. 
Top: Heat map of $pn$ knockout events as a function of the opening angle $\cos 
\gamma$ and the initial neutron momentum $p_2$. We compare measured events (left 
panel adapted from Ref.~\cite{Piasetzky:2006ai}) with calculated cross sections 
(right panel).
Bottom: The cross section as a function of the light-cone momentum fractions 
$\alpha_m$ and $\alpha_2$. We compare calculated cross sections (solid lines), 
to the histogram of the measured number of events~\cite{Tang:2002ww}. As the 
data are provided as ``counts'' they} are scaled in such a way that the mean of 
the two  measured and computed 
cross-section peaks coincide.
\label{fig_pop}
\end{figure}

In order to test the validity of the factorized cross section derived in 
Sec.~\ref{sed:formalism}, we compare computed  
${}^{\,12}\text{C}(p,p^{\prime}pn)$ cross sections 
with experimental results obtained with the EVA spectrometer at 
BNL~\cite{Tang:2002ww,Piasetzky:2006ai}. The kinematics is detailed in 
Ref.~\cite{Malki:2000gh}. In order to match the kinematic areas of the 
calculations and the data we have adopted the kinematics cuts outlined in 
Ref.~\cite{Tang:2002ww} and we used the data-driven $pp$ differential cross 
section parametrization of Ref.~\protect\cite{Uzhinsky:2016uhg}. We consider 
knockout of SRC $pn$ pairs from the $s$ and 
$p$ shells in carbon. All results of Fig.~\ref{fig_pop} use 
Eq.~(\ref{eqCrossSection}) with the ejected neutron as the ``slow'' nucleon. 

Obviously the calculations reproduce the change in the angular correlations for 
the opening angle between the nucleons in the pair between $p_2 < k_F $  and 
$p_2 > k_F$. 
In Fig.~\ref{fig_pop}, we also show the ${}^{\,12}\text{C}(p,p^{\prime}pn)$ 
cross section as a function of the longitudinal light-cone momentum fractions 
$\alpha_m$ and $\alpha_2$ carried by the nucleons in the ejected SRC NN pair. 
With the $\hat{z}$-axis parallel to the beam, the $\alpha_j$ are defined as 
\cite{Frankfurt:1988nt} 
\begin{align}
\alpha_j = A \, \frac{E_j - p_{j,z}}{E_A - p_{A,z}} \, .
\end{align}
For high-energetic beams, the $\alpha_j$ are natural variables to characterize 
the nucleons' momentum distributions. In the PF ($P_A = (m_A, \vec{0})$) 
one finds~\cite{Tang:2002ww}
\begin{align}
\label{eqApproxAlphas}
\alpha_m \approx 1 + \frac{E_m - p_{m,z}}{m_{N_1}}
\qquad \qquad \text{and} \qquad \qquad
\alpha_2 \approx \frac{E_2 - p_{2,z}}{m_{N_2}} \, .
\end{align}

In Fig.~\ref{fig_pop} we compare the calculated cross sections with the measured 
number of events. As a supplementary consistency check, we apply the same 
scaling factor to the $\alpha_m$ and $\alpha_2$ distributions.
The shapes of both histograms are captured  well by the calculated cross 
sections. The use of a unique normalization factor 
indicates that our model reproduces the relative cross sections $ \frac {d 
\sigma} {d \alpha_m} / \frac {d \sigma} {d \alpha_2} $.

In order to derive the SRC pair probabilities from the cross section of 
Eq.~(\ref{eqCrossSection}), we define the function 
\begin{align}
\label{eqRemovalProb}
P^{M_T}_{\beta \gamma}(\vec{P}, \vec{k})
=
 2^{2 \left\vert M_T \right\vert - 1} \;
 \sum_{J \; M \; T}
 \frac{1}{2 J + 1}
 \frac{1}{2 T + 1}
 F^{\beta \gamma}_{JM,T} (\vec{P}, \vec{k}) \, ,
\end{align}
that determines at given c.m. and relative momentum the cross section for 
removing a pair with given $M_T$. The ratio 
of $pn$ to $pp$ SRC pair probabilities at given  $k$ can be derived from the 
expressions 
\ref{eq:Ffunction}, \ref{eqLinearIntegralTransformation},  
\ref{eq:roundPfunction} after integration over the c.m.~momentum $\vec{P}$ of 
the SRC pair. This operation results in 
\begin{align}
\label{eqCrossSectionRatio}
 \mathcal{R}_{\frac {pn}{pp}}(k)
=
 \frac{ \sum _{\beta \gamma}
 \int \text{d} \Omega_{k}
\text{d} \vec{P} \;
 P^{M_T=0}_{\beta \gamma}(\vec{P}, \vec{k})
 }{ \sum _{\beta \gamma}
 \int \text{d} \Omega_{k}
 \text{d} \vec{P} \;
 P^{M_T=1}_{\beta \gamma}(\vec{P}, \vec{k})}
=
 \frac{1}{2}
 +
 \frac{3}{2} \;
 \mathcal{S}_0
 +
 108 \;
 \mathcal{S}_1 
 \left\lbrace
 \frac{
 \mathcal{F}^{(2)}_\nu [f_{t \tau}] (k)
 }
 {
 \mathcal{F}^{(0)}_{\nu} [f_c - 3 f_{\sigma \tau}] (k)
 }
 \right\rbrace^2 \, ,
\end{align}
where $\mathcal{S}_0$ and $\mathcal{S}_1$ depend on the quantum numbers of the 
impinging ion and are $k$-independent. The ratio of 
Eq.~(\ref{eqCrossSectionRatio}) can be connected to  cross-section ratios 
accessible in $p(A,p^{\prime}pN)$ experiments and is for example not 
sensitive to FSI effects in the SRC pair.

\begin{figure}
\centering
\includegraphics[width=\textwidth]{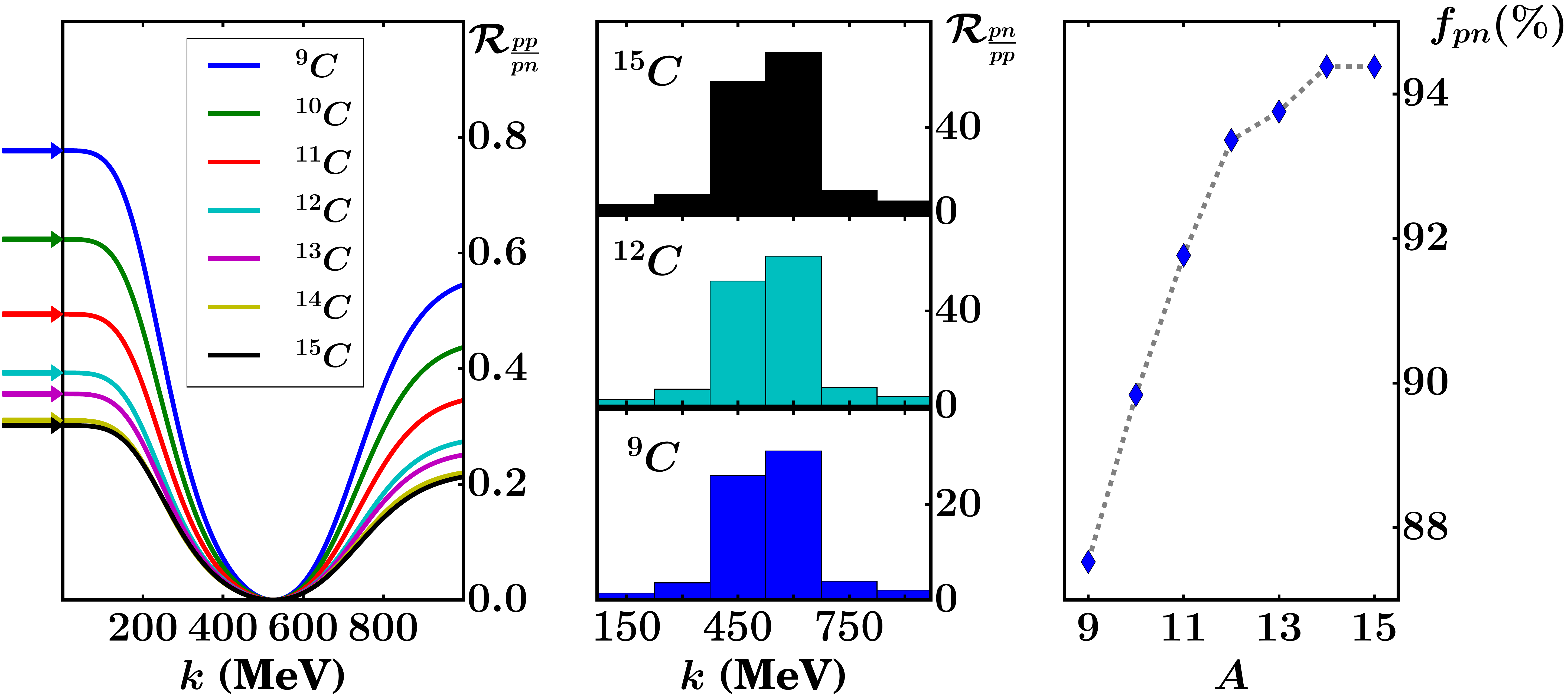} 
\caption{Ratios of the $pn$ and $pp$ SRC-pair removal probabilities in various carbon isotopes. Left: the $pp$ over $pn$ ratio, this is the reciprocal of Eq.~(\ref{eqCrossSectionRatio}), as a function of the initial relative momentum $k$ of the SRC NN pair. 
The arrows indicate the constant ratio $\frac{1}{2}+\frac{3}{2}S_0$ when no tensor correlations are taken into account.
Center: the $k$ dependence of the binned $pn$ over $pp$ ratios of Eq.~(\ref{eqCrossSectionRatio}). Right: the fraction of SRC $pn$ pairs defined in Eq.~(\ref{eqFraction}) for different carbon isotopes.}
\label{figCrossSectionRatio}
\end{figure}

Fig.~\ref{figCrossSectionRatio} shows predictions for the $k$-dependence of the 
ratio (\ref{eqCrossSectionRatio}) and its inverse for different carbon 
isotopes. Three regions in the relative momentum can be discerned. For $k<k_F$ 
the $pp$ to $pn$ SRC pair fractions are constant and are determined by 
$\left(\mathcal{R}_{\frac{pn}{pp}}(k) = \frac{1}{2}+\frac{3}{2}S_0\right)$. The 
number of $n=0,\; l=0$ $pn$ and $pp$ pairs determines the value of the ratio. 
 A rapid decrease of the $pp/pn$ ratio is seen to start at $k \gtrsim k_F$. For 
all considered carbon isotopes the SRC $pn$ removal probability is much larger 
than the $pp$ one for $k_F \lesssim k \lesssim 3 k_F$. The dominance of the  
$pn$ SRC pairs over the $pp$ SRC ones in that momentum range is clearly 
illustrated in the center panel of Fig.~\ref{figCrossSectionRatio}. For all 
carbon isotopes the SRC-driven $p(\text{C},p^{\prime}pn)$ cross section is at 
least an order of magnitude larger than the $p(\text{C},p^{\prime}pp)$ one. We 
can conclude that for all isotopes, $pn$ pair knockout dominates for $k_F 
\lesssim k \lesssim 3 k_F$, a property that is  observed experimentally in 
electro-induced nucleon pair knockout from 
$^{12}$C~\cite{Hen:2014nza,Subedi:2008zz}. For $k \gtrsim 3 k_F$ the cross 
sections are no longer dominated by the tensor correlation function, and the 
$pp$ over $pn$ SRC-pair probabilities display similar trends as observed for $k 
\lesssim k_F$. 
The expected fraction of $pn$ over $pp$ SRC pair removals for $k_F \lesssim k \lesssim 3 k_F$ can be defined as  
\begin{align}
\label{eqFraction}
 f_{pn} 
=  
 \frac{ \sum _{\beta \gamma}
 \int_{k_F}^{3 k_F} \text{d} k
 \int \text{d} \Omega_{k}
 \int \text{d} \vec{P} \;
 P^{0}_{\beta \gamma} (\vec{P}, \vec{k})
 }{ \sum_{\beta \gamma}
 \int_{k_F}^{3 k_F} \text{d} k
 \int \text{d} \Omega_{k}
 \int \text{d} \vec{P} \;
 \left[ P^{1}_{\beta \gamma} (\vec{P}, \vec{k}) +
 P^{0}_{\beta \gamma} (\vec{P}, \vec{k}) \right] \; 
 } \; .
\end{align}
In Fig.~\ref{figCrossSectionRatio} we display predictions for $f_{pn}$ as a function of $A$ for $Z=6$. The observed trends can be mainly attributed to the number of $(n=0,\;l=0)$ $pn$ and $pp$ pairs for given $\left(N,Z \right)$. For $^{12}$C, we find a fraction $\approx 93 \%$ that is comparable to the fractions extracted from measurements~\cite{Subedi:2008zz, Hen:2014nza}. Our predictions for $f_{pn}$ cover neutron excess values $(N-Z)/Z$ between -0.5 and +0.5. Over that range the $f_{pn}$ changes by less than $10\%$, a value much smaller than anticipated on the basis of variations in the ratio $(N\times Z)$/$(Z \times Z)$. 

\section{Conclusions}
\label{sec:concl}

We have introduced a factorized plane-wave-based model for the calculation of SRC-driven $p(A,p^\prime N_1 N_2 R^\ast)$ reaction cross sections, that can serve in particular for the analysis of exclusive proton-induced two-nucleon knockout reactions in inverse kinematics. Our model calculates these cross sections based on a chosen set of single-particle wave functions, correlation functions and  free $pN$-scattering cross section data. We have shown that our model reproduces characteristic features of SRC-driven two-nucleon knockout reactions that are also found in electro-induced two-nucleon knockout reactions. We also describe  ${}^{\,12}\text{C}(p,p^{\prime}pn)$ data rather well. 
{Based on the factorization properties of the cross section, we can conclude that the isospin dependence of SRC can be studied by evaluating  cross-section ratios. The ratio of the integrated removal probabilities for $pn$ over $pp$ pairs can be connected to a ratio of correlation functions depending on the initial relative momentum of the pair.}
Our model is an important first step in constructing a reaction model. 
Required refinements to the proposed model include the description of IFSI 
and the inclusion of configuration-mixing effects  \cite{ISI:000385248700002}  
in the description of the ground-state wave function of the target nuclei.  
The model is applicable to planned experiments aimed at further uncovering the 
characteristics of nuclear SRC, in particular its asymmetry dependence. 

\section*{Acknowledgements}
The computational resources (Stevin Supercomputer Infrastructure) and services used in this work were provided by the VSC (Flemish Supercomputer Center), funded by Ghent University, FWO and the Flemish Government – department EWI

\section*{References}
\bibliography{bibliography.bib}

\begin{thebibliography}{10}
\expandafter\ifx\csname url\endcsname\relax
  \def\url#1{\texttt{#1}}\fi
\expandafter\ifx\csname urlprefix\endcsname\relax\def\urlprefix{URL }\fi
\expandafter\ifx\csname href\endcsname\relax
  \def\href#1#2{#2} \def\path#1{#1}\fi

\bibitem{Arrington:2011xs}
J.~Arrington, D.~W. Higinbotham, G.~Rosner, M.~Sargsian, {Hard probes of
  short-range nucleon-nucleon correlations}, Prog. Part. Nucl. Phys. 67 (2012)
  898--938.
\newblock \href {http://arxiv.org/abs/1104.1196} {\path{arXiv:1104.1196}},
  \href {http://dx.doi.org/10.1016/j.ppnp.2012.04.002}
  {\path{doi:10.1016/j.ppnp.2012.04.002}}.

\bibitem{Atti:2015eda}
C.~Ciofi~degli Atti, {In-medium short-range dynamics of nucleons: Recent
  theoretical and experimental advances}, Phys. Rept. 590 (2015) 1--85.
\newblock \href {http://dx.doi.org/10.1016/j.physrep.2015.06.002}
  {\path{doi:10.1016/j.physrep.2015.06.002}}.

\bibitem{Hen:2016kwk}
O.~Hen, G.~A. Miller, E.~Piasetzky, L.~B. Weinstein, {Nucleon-Nucleon
  Correlations, Short-lived Excitations, and the Quarks Within}\href
  {http://arxiv.org/abs/1611.09748} {\path{arXiv:1611.09748}}.

\bibitem{Dickhoff:2004xx}
W.~H. Dickhoff, C.~Barbieri, {Selfconsistent Green's function method for nuclei
  and nuclear matter}, Prog. Part. Nucl. Phys. 52 (2004) 377--496.
\newblock \href {http://arxiv.org/abs/nucl-th/0402034}
  {\path{arXiv:nucl-th/0402034}}, \href
  {http://dx.doi.org/10.1016/j.ppnp.2004.02.038}
  {\path{doi:10.1016/j.ppnp.2004.02.038}}.

\bibitem{Hen:2014nza}
O.~Hen, et~al., {Momentum sharing in imbalanced Fermi systems}, Science 346
  (2014) 614--617.
\newblock \href {http://arxiv.org/abs/1412.0138} {\path{arXiv:1412.0138}},
  \href {http://dx.doi.org/10.1126/science.1256785}
  {\path{doi:10.1126/science.1256785}}.

\bibitem{Wiringa:2013ala}
R.~B. Wiringa, R.~Schiavilla, S.~C. Pieper, J.~Carlson, {Nucleon and
  nucleon-pair momentum distributions in $A \le 12$ nuclei}, Phys. Rev. C89~(2)
  (2014) 024305.
\newblock \href {http://arxiv.org/abs/1309.3794} {\path{arXiv:1309.3794}},
  \href {http://dx.doi.org/10.1103/PhysRevC.89.024305}
  {\path{doi:10.1103/PhysRevC.89.024305}}.

\bibitem{Vanhalst:2014cqa}
J.~Ryckebusch, W.~Cosyn, M.~Vanhalst, {Stylized features of single-nucleon
  momentum distributions}, J. Phys. G42~(5) (2015) 055104.
\newblock \href {http://arxiv.org/abs/1405.3814} {\path{arXiv:1405.3814}},
  \href {http://dx.doi.org/10.1088/0954-3899/42/5/055104}
  {\path{doi:10.1088/0954-3899/42/5/055104}}.

\bibitem{Fortin:2016hny}
M.~Fortin, C.~Providencia, A.~R. Raduta, F.~Gulminelli, J.~L. Zdunik,
  P.~Haensel, M.~Bejger, {Neutron star radii and crusts: uncertainties and
  unified equations of state}, Phys. Rev. C94~(3) (2016) 035804.
\newblock \href {http://arxiv.org/abs/1604.01944} {\path{arXiv:1604.01944}},
  \href {http://dx.doi.org/10.1103/PhysRevC.94.035804}
  {\path{doi:10.1103/PhysRevC.94.035804}}.

\bibitem{Oertel:2016bki}
M.~Oertel, M.~Hempel, T.~Klähn, S.~Typel, {Equations of state for supernovae
  and compact stars}, Rev. Mod. Phys. 89~(1) (2017) 015007.
\newblock \href {http://arxiv.org/abs/1610.03361} {\path{arXiv:1610.03361}},
  \href {http://dx.doi.org/10.1103/RevModPhys.89.015007}
  {\path{doi:10.1103/RevModPhys.89.015007}}.

\bibitem{Onderwater:1997zz}
C.~J.~G. Onderwater, et~al., {Dominance of S-01 Proton-Pair Emission in the
  O-16 (e, e-prime pp) Reaction}, Phys. Rev. Lett. 78 (1997) 4893--4897.
\newblock \href {http://dx.doi.org/10.1103/PhysRevLett.78.4893}
  {\path{doi:10.1103/PhysRevLett.78.4893}}.

\bibitem{Giusti:1997pa}
C.~Giusti, F.~D. Pacati, K.~Allaart, W.~J.~W. Geurts, W.~H. Dickhoff,
  H.~Muther, {Selectivity of the O-16(e,e-prime p p) reaction to discrete final
  states}, Phys. Rev. C57 (1998) 1691--1702.
\newblock \href {http://arxiv.org/abs/nucl-th/9709021}
  {\path{arXiv:nucl-th/9709021}}, \href
  {http://dx.doi.org/10.1103/PhysRevC.57.1691}
  {\path{doi:10.1103/PhysRevC.57.1691}}.

\bibitem{Onderwater:1998zz}
C.~J.~G. Onderwater, et~al., {Signatures for Short-Range Correlations in O-16
  Observed in the Reaction O-16 (e, e-prime pp) C-14}, Phys. Rev. Lett. 81
  (1998) 2213--2216.
\newblock \href {http://dx.doi.org/10.1103/PhysRevLett.81.2213}
  {\path{doi:10.1103/PhysRevLett.81.2213}}.

\bibitem{ROSNER200099}
G.~Rosner,
  \href{http://www.sciencedirect.com/science/article/pii/S0146641000000636}{Probing
  short-range nucleon-nucleon correlations with virtual photons at mami},
  Progress in Particle and Nuclear Physics 44 (2000) 99 -- 112.
\newblock \href
  {http://dx.doi.org/http://dx.doi.org/10.1016/S0146-6410(00)00063-6}
  {\path{doi:http://dx.doi.org/10.1016/S0146-6410(00)00063-6}}.
\newline\urlprefix\url{http://www.sciencedirect.com/science/article/pii/S0146641000000636}

\bibitem{Starink:2000qhh}
R.~Starink, et~al., {Evidence for short-range correlations in 16 O}, Phys.
  Lett. B474 (2000) 33--40.
\newblock \href {http://dx.doi.org/10.1016/S0370-2693(99)01510-5}
  {\path{doi:10.1016/S0370-2693(99)01510-5}}.

\bibitem{Tang:2002ww}
A.~Tang, et~al., {n-p short range correlations from (p,2p + n) measurements},
  Phys. Rev. Lett. 90 (2003) 042301.
\newblock \href {http://arxiv.org/abs/nucl-ex/0206003}
  {\path{arXiv:nucl-ex/0206003}}, \href
  {http://dx.doi.org/10.1103/PhysRevLett.90.042301}
  {\path{doi:10.1103/PhysRevLett.90.042301}}.

\bibitem{Shneor:2007tu}
R.~Shneor, et~al., {Investigation of proton-proton short-range correlations via
  the C-12(e, e-prime pp) reaction}, Phys. Rev. Lett. 99 (2007) 072501.
\newblock \href {http://arxiv.org/abs/nucl-ex/0703023}
  {\path{arXiv:nucl-ex/0703023}}, \href
  {http://dx.doi.org/10.1103/PhysRevLett.99.072501}
  {\path{doi:10.1103/PhysRevLett.99.072501}}.

\bibitem{Korover:2014dma}
I.~Korover, et~al., {Probing the Repulsive Core of the Nucleon-Nucleon
  Interaction via the $^4He(e,e′pN)$ Triple-Coincidence Reaction}, Phys. Rev.
  Lett. 113~(2) (2014) 022501.
\newblock \href {http://arxiv.org/abs/1401.6138} {\path{arXiv:1401.6138}},
  \href {http://dx.doi.org/10.1103/PhysRevLett.113.022501}
  {\path{doi:10.1103/PhysRevLett.113.022501}}.

\bibitem{Colle:2015ena}
C.~Colle, O.~Hen, W.~Cosyn, I.~Korover, E.~Piasetzky, J.~Ryckebusch, L.~B.
  Weinstein, {Extracting the mass dependence and quantum numbers of short-range
  correlated pairs from $A(e,e′p)$ and $A(e,e′pp)$ scattering}, Phys. Rev.
  C92~(2) (2015) 024604.
\newblock \href {http://arxiv.org/abs/1503.06050} {\path{arXiv:1503.06050}},
  \href {http://dx.doi.org/10.1103/PhysRevC.92.024604}
  {\path{doi:10.1103/PhysRevC.92.024604}}.

\bibitem{Ryckebusch:1996wc}
J.~Ryckebusch, {Photoinduced two proton knockout and ground state correlations
  in nuclei}, Phys. Lett. B383 (1996) 1--8.
\newblock \href {http://arxiv.org/abs/nucl-th/9605043}
  {\path{arXiv:nucl-th/9605043}}, \href
  {http://dx.doi.org/10.1016/0370-2693(96)00725-3}
  {\path{doi:10.1016/0370-2693(96)00725-3}}.

\bibitem{Colle:2015lyl}
C.~Colle, W.~Cosyn, J.~Ryckebusch, {Final-state interactions in two-nucleon
  knockout reactions}, Phys. Rev. C93~(3) (2016) 034608.
\newblock \href {http://arxiv.org/abs/1512.07841} {\path{arXiv:1512.07841}},
  \href {http://dx.doi.org/10.1103/PhysRevC.93.034608}
  {\path{doi:10.1103/PhysRevC.93.034608}}.

\bibitem{Aumann20073}
T.~Aumann,
  \href{http://www.sciencedirect.com/science/article/pii/S0146641006000974}{Prospects
  of nuclear structure at the future \{GSI\} accelerators}, Progress in
  Particle and Nuclear Physics 59~(1) (2007) 3 -- 21, international Workshop on
  Nuclear Physics 28th CourseRadioactive Beams, Nuclear Dynamics and
  AstrophysicsEttore Majorana Center for Scientific Culture.
\newblock \href {http://dx.doi.org/https://doi.org/10.1016/j.ppnp.2006.12.018}
  {\path{doi:https://doi.org/10.1016/j.ppnp.2006.12.018}}.
\newline\urlprefix\url{http://www.sciencedirect.com/science/article/pii/S0146641006000974}

\bibitem{Kobayashi:2008zzc}
T.~Kobayashi, et~al., {(p,2p) reactions on C-(9-16) at 250-MeV/A}, Nucl. Phys.
  A805 (2008) 431--438.
\newblock \href {http://dx.doi.org/10.1016/j.nuclphysa.2008.02.282}
  {\path{doi:10.1016/j.nuclphysa.2008.02.282}}.

\bibitem{Panin:2016div}
V.~Panin, et~al., {Exclusive measurements of quasi-free proton scattering
  reactions in inverse and complete kinematics}, Phys. Lett. B753 (2016)
  204--210.
\newblock \href {http://dx.doi.org/10.1016/j.physletb.2015.11.082}
  {\path{doi:10.1016/j.physletb.2015.11.082}}.

\bibitem{Jacob:1973tu}
G.~Jacob, T.~A.~J. Maris, {Quasi-free scattering and nuclear structure. 2.},
  Rev. Mod. Phys. 45 (1973) 6--21.
\newblock \href {http://dx.doi.org/10.1103/RevModPhys.45.6}
  {\path{doi:10.1103/RevModPhys.45.6}}.

\bibitem{Kitching:1984gc}
P.~Kitching, W.~J. McDonald, T.~A.~J. Maris, C.~A.~Z. Vasconcellos, {Recent
  Developments in Quasifree Nucleon-nucleon Scattering}, Adv. Nucl. Phys. 15
  (1985) 43--83.

\bibitem{Ryckebusch:1997gn}
J.~Ryckebusch, V.~Van~der Sluys, K.~Heyde, H.~Holvoet, W.~Van~Nespen,
  M.~Waroquier, M.~Vanderhaeghen, {Electroinduced two nucleon knockout and
  correlations in nuclei}, Nucl. Phys. A624 (1997) 581--622.
\newblock \href {http://arxiv.org/abs/nucl-th/9702049}
  {\path{arXiv:nucl-th/9702049}}, \href
  {http://dx.doi.org/10.1016/S0375-9474(97)00385-0}
  {\path{doi:10.1016/S0375-9474(97)00385-0}}.

\bibitem{Blomqvist:1998gq}
K.~I. Blomqvist, et~al., {Investigation of short range nucleon nucleon
  correlations using the reaction C-12(e,e' p p) in close to 4pi geometry},
  Phys. Lett. B421 (1998) 71--78.
\newblock \href {http://dx.doi.org/10.1016/S0370-2693(98)00024-0}
  {\path{doi:10.1016/S0370-2693(98)00024-0}}.

\bibitem{Colle:2013nna}
C.~Colle, W.~Cosyn, J.~Ryckebusch, M.~Vanhalst, {Factorization of exclusive
  electron-induced two-nucleon knockout}, Phys. Rev. C89~(2) (2014) 024603.
\newblock \href {http://arxiv.org/abs/1311.1980} {\path{arXiv:1311.1980}},
  \href {http://dx.doi.org/10.1103/PhysRevC.89.024603}
  {\path{doi:10.1103/PhysRevC.89.024603}}.

\bibitem{Stetz:1980zz}
A.~W. Stetz, {Analysis of quasi-free scattering data in the impulse
  approximation}, Phys. Rev. C21 (1980) 1979--1988.
\newblock \href {http://dx.doi.org/10.1103/PhysRevC.21.1979}
  {\path{doi:10.1103/PhysRevC.21.1979}}.

\bibitem{Guardiola:1996dq}
R.~Guardiola, P.~I. Moliner, J.~Navarro, R.~F. Bishop, A.~Puente, N.~R. Walet,
  {Translationally invariant treatment of pair correlations in nuclei. 1. Spin
  and isospin dependent correlations}, Nucl. Phys. A609 (1996) 218--236.
\newblock \href {http://arxiv.org/abs/nucl-th/9607024}
  {\path{arXiv:nucl-th/9607024}}, \href
  {http://dx.doi.org/10.1016/0375-9474(96)00315-6}
  {\path{doi:10.1016/0375-9474(96)00315-6}}.

\bibitem{Benhar:1993zz}
O.~Benhar, V.~R. Pandharipande, S.~C. Pieper, {Electron-scattering studies of
  correlations in nuclei}, Rev. Mod. Phys. 65 (1993) 817--828.
\newblock \href {http://dx.doi.org/10.1103/RevModPhys.65.817}
  {\path{doi:10.1103/RevModPhys.65.817}}.

\bibitem{Pieper:1992gr}
S.~C. Pieper, R.~B. Wiringa, V.~R. Pandharipande, {Variational calculation of
  the ground state of O-16}, Phys. Rev. C46 (1992) 1741--1756.
\newblock \href {http://dx.doi.org/10.1103/PhysRevC.46.1741}
  {\path{doi:10.1103/PhysRevC.46.1741}}.

\bibitem{CiofidegliAtti:1999zz}
C.~Ciofi~degli Atti, D.~Treleani, {Linked cluster expansion for the calculation
  of the semi-inclusive A (e, e-prime p) X processes using correlated Glauber
  wave functions}, Phys. Rev. C60 (1999) 024602.
\newblock \href {http://dx.doi.org/10.1103/PhysRevC.60.024602}
  {\path{doi:10.1103/PhysRevC.60.024602}}.

\bibitem{Vanhalst:2012ur}
M.~Vanhalst, J.~Ryckebusch, W.~Cosyn, {Quantifying short-range correlations in
  nuclei}, Phys. Rev. C86 (2012) 044619.
\newblock \href {http://arxiv.org/abs/1206.5151} {\path{arXiv:1206.5151}},
  \href {http://dx.doi.org/10.1103/PhysRevC.86.044619}
  {\path{doi:10.1103/PhysRevC.86.044619}}.

\bibitem{Gearheart1994}
C.~Gearheart, Ph.D. thesis, Washington University, St. Louis (1994).

\bibitem{Arndt:2000xc}
R.~A. Arndt, I.~I. Strakovsky, R.~L. Workman, {Nucleon nucleon elastic
  scattering to 3 GeV}, Phys. Rev. C62 (2000) 034005.
\newblock \href {http://arxiv.org/abs/nucl-th/0004039}
  {\path{arXiv:nucl-th/0004039}}, \href
  {http://dx.doi.org/10.1103/PhysRevC.62.034005}
  {\path{doi:10.1103/PhysRevC.62.034005}}.

\bibitem{ScatteringSAID}
{Scattering Analysis Interactive Dial-in Program (SAID)},
  \url{http://gwdac.phys.gwu.edu/}.

\bibitem{Uzhinsky:2016uhg}
V.~Uzhinsky, A.~Galoyan, Q.~Hu, J.~Ritman, H.~Xu, {Empirical parametrization of
  the nucleon-nucleon elastic scattering amplitude at high beam momenta for
  Glauber calculations and Monte Carlo simulations}, Phys. Rev. C94~(6) (2016)
  064003.
\newblock \href {http://arxiv.org/abs/1603.04731} {\path{arXiv:1603.04731}},
  \href {http://dx.doi.org/10.1103/PhysRevC.94.064003}
  {\path{doi:10.1103/PhysRevC.94.064003}}.

\bibitem{VanCuyck:2016fab}
T.~Van~Cuyck, N.~Jachowicz, R.~González-Jiménez, M.~Martini, V.~Pandey,
  J.~Ryckebusch, N.~Van~Dessel, {Influence of short-range correlations in
  neutrino-nucleus scattering}, Phys. Rev. C94~(2) (2016) 024611.
\newblock \href {http://arxiv.org/abs/1606.00273} {\path{arXiv:1606.00273}},
  \href {http://dx.doi.org/10.1103/PhysRevC.94.024611}
  {\path{doi:10.1103/PhysRevC.94.024611}}.

\bibitem{Piasetzky:2006ai}
E.~Piasetzky, M.~Sargsian, L.~Frankfurt, M.~Strikman, J.~W. Watson, {Evidence
  for the strong dominance of proton-neutron correlations in nuclei}, Phys.
  Rev. Lett. 97 (2006) 162504.
\newblock \href {http://arxiv.org/abs/nucl-th/0604012}
  {\path{arXiv:nucl-th/0604012}}, \href
  {http://dx.doi.org/10.1103/PhysRevLett.97.162504}
  {\path{doi:10.1103/PhysRevLett.97.162504}}.

\bibitem{Malki:2000gh}
A.~Malki, et~al., {Backward emitted high-energy neutrons in hard reactions of p
  and pi+ on carbon}, Phys. Rev. C65 (2002) 015207.
\newblock \href {http://arxiv.org/abs/nucl-ex/0005006}
  {\path{arXiv:nucl-ex/0005006}}, \href
  {http://dx.doi.org/10.1103/PhysRevC.65.015207}
  {\path{doi:10.1103/PhysRevC.65.015207}}.

\bibitem{Frankfurt:1988nt}
L.~L. Frankfurt, M.~I. Strikman, {Hard Nuclear Processes and Microscopic
  Nuclear Structure}, Phys. Rept. 160 (1988) 235--427.
\newblock \href {http://dx.doi.org/10.1016/0370-1573(88)90179-2}
  {\path{doi:10.1016/0370-1573(88)90179-2}}.

\bibitem{Subedi:2008zz}
R.~Subedi, et~al., {Probing Cold Dense Nuclear Matter}, Science 320 (2008)
  1476--1478.
\newblock \href {http://arxiv.org/abs/0908.1514} {\path{arXiv:0908.1514}},
  \href {http://dx.doi.org/10.1126/science.1156675}
  {\path{doi:10.1126/science.1156675}}.

\end{thebibliography}

\end{document}